______________________________________________________________________
\documentstyle[eqsecnum,aps]{revtex}
\begin{document}

\title{Irreducible bases in icosahedral group space}
 
\author{Shi-Hai Dong}

\address{Institute of High Energy Physics, P.O. Box 918(4), 
Beijing 100039, The People's Republic of China }

\author{Xi-Wen Hou}

\address{Institute of High Energy Physics, 
P.O. Box 918(4), Beijing 100039,\\
and Department of Physics, University of Three Gorges, 
Yichang 443000, The People's Republic of China }

\author{Mi Xie}

\address{Department of Physics, Tianjin Normal University, 
Tianjin 300074}

\author{Zhong-Qi Ma \thanks{Electronic address: MAZQ@BEPC3.IHEP.AC.CN}}

\address{Institute of High Energy Physics, 
P.O.Box 918(4), Beijing 100039, The People's Republic of China}

\maketitle

\vspace{14mm}
\begin{abstract}
The irreducible bases in the icosahedral group space are 
calculated explicitly by reducing the regular representation. 
The symmetry adapted bases of the system with {\bf I} or 
{\bf I}$_{h}$ symmetry can be calculated easily and generally 
by applying those irreducible bases to wavefunctions of the 
system, if they are not vanishing. As examples, the submatrices 
of the H\"{u}ckel Hamiltonians for Carbon-60 and Carbon-240 
are re-calculated by the irreducible bases.

\vspace{4mm}


\vspace{4mm}

\noindent
Running head:~~~~Irreducible bases in icosahedral group space
 
\end{abstract}


\newpage
\noindent
{\bf I. INTRODUCTION}

\vspace{3mm}
Fullerences (Kroto 1988, Huffman 1991, Pennis 1991), 
such as $B_{12}H_{12}$, 
$C_{20}H_{20}$ and $C_{60}$, are intriguing cage-like 
molecules of carbon atoms with the icosahedral symmetry.
This discovery (Rohlfing {\it et al}. 1984, Kroto {\it et al}. 1985,
Weeks and Harter 1989) has greatly drawn 
attentions of chemists and physicists (Deng and Yang 1992, 
Chou and Yang 1993, Friedberg and Lee 1992). 
With the development of experimental techniques 
in high resolution spectroscopy, many new data on vibrational 
spectra of polyatomic molecules with the symmetry {\bf I}$_{h}$ 
were observed and analyzed (Negri and Orlandi 1996, Olthof
{\it et al}. 1996, Giannozzi and Baroni 1994, Schettino {\it et al}.
1994, Gunnarsson {\it et al}. 1995, Doye and Wales 1996, 
Wang {\it et al}. 1996, Tang {\it et al}. 1996, Tang and Huang
1997). The vibrational modes, the force fields, and the spin-orbit 
coupling coefficients for icosahedral molecules were studied 
in some detail (Clougherty and Gorman 1996, Mart\'{i}nez-Torres
{\it et al}. 1996, Varga {\it et al}. 1996, Fowler and Ceulemans 1985).

As is well known, 
symmetry analysis provides a powerful tool for classifying 
energy levels and organizing experimental data. In explaining 
the vibrational spectra of polyatomic molecules, the symmetry 
adapted bases (SAB) play an important role in simplifying the 
calculations (Lemus and Frank 1994, Ma {\it et al}. 1996, 
Chen {\it et al}. 1996). The SAB are defined as the 
orthogonal bases that belong to given rows of given irreducible 
representations of the symmetry group. In those studies, 
SAB of the system with the {\bf I}$_{h}$ symmetry were
widely used. Therefore, the properties of the {\bf I}$_{h}$
group are worthy to be studied in some detail, although the 
dimension of the {\bf I}$_{h}$ group space is 120.

Early works on {\bf I} group are mainly concerned with the 
construction of the representations of {\bf I} subduced by 
$D^{\ell}$ of SO(3) group and the 3j- and 6j-symbols 
(McLellan 1961, Golding 1973, Pooler 1980, Brown 1987). 
Liu-Ping-Chen (1990) enumerated the 60 elements of 
icosahedral group {\bf I}, listed its group table, and 
calculated the irreducible representation matrices of all the 
60 elements explicitly. The character tables of the point
groups and the double point groups were listed (Altmann and Herzig 1994,
Balasubramanian 1996). Recently, Chen and Ping (1997) constructed 
the point-group symmetrized boson representation, 
and gave the explicit expressions of the SAB for seven 
important cases of the molecule $B_{12}H_{12}$.  

As another approach, in this paper we will explicitly 
calculate the irreducible bases $\psi_{\mu \nu}^{\Gamma}$
in the group spaces of {\bf I} and {\bf I}$_{h}$ by reducing 
the regular representation of {\bf I}:
$$R\psi_{\mu \nu}^{\Gamma}=\displaystyle \sum_{\rho}~
\psi_{\rho \nu}^{\Gamma} D_{\rho \mu}^{\Gamma}(R),~~~~
\psi_{\mu \nu}^{\Gamma}R=\displaystyle \sum_{\rho}~
D_{\nu \rho}^{\Gamma}(R)\psi_{\mu \rho}^{\Gamma},~~~~
R\in {\bf I}. (1) $$

\noindent
where $D^{\Gamma}$ is an irreducible representation of
{\bf I}, and $\psi_{\mu \nu}^{\Gamma}$ is a combination 
of the group elements. Applying those irreducible bases 
to any function $F(x)$, if it is not vanishing, one will 
obtain the SAB $\psi_{\mu \nu}^{\Gamma}F(x)$:
$$R\left\{\psi_{\mu \nu}^{\Gamma}F(x)\right\}=\displaystyle \sum_{\rho}~
\left\{\psi_{\rho \nu}^{\Gamma}F(x)\right\} D_{\rho \mu}^{\Gamma}(R).
\eqno (2) $$

\noindent
It is an unified and straightforward way to calculate the SAB
of the system with the {\bf I}$_{h}$ symmetry. 

By the way, we would like to point out that the rank of group 
{\bf I} is two, not three (McLellan 1961, Liu {\it et al}. 1990,
Lomont 1959). It means that all 60 
elements of {\bf I} can be expressed as the products of only 
two generators. 

The plan of this paper is as follows. In Sec. II we will give our 
notations. In Sec. III the irreducible bases 
in the {\bf I} group space are calculated explicitly, and the 
irreducible bases of {\bf I}$_{h}$ are easy to be calculated 
from those of {\bf I}. Three examples are given to explain 
how to calculate the SAB in terms of those irreducible bases. 
A short conclusion is given in Sec. IV.

\vspace{5mm}
\noindent
{\bf II. NOTATIONS AND GENERATORS OF GROUP {\bf I}}

\vspace{3mm}
A regular icosahedron is shown in Fig.1. The vertices on the upper
part are labeled by $A_{j}$, $0\leq j \leq 5$, and their opposite 
vertices by $B_{j}$. The $z$ and $y$ axes point from the center
$O$ to $A_{0}$ and the midpoint of $A_{2}B_{5}$, respectively.

\begin{center}
\fbox{Fig. 1.}
\end{center}

The group {\bf I} has 6 five-fold axes, 10 three-fold axes, and 15 
two-fold axes. One of the five-fold axes directs along $z$ axis, and
the rest point from $B_{j}$ to $A_{j}$ ($1 \leq j \leq 5$) with the
polar angle $\theta_{1}$ and azimuthal angles $\varphi_{j}^{(1)}$. 
The rotations through $2\pi/5$ around those five-fold axes are denoted by 
$T_{j}$, $0\leq j \leq 5$. The three-fold axes join the centers of two 
opposite faces. The polar angles of the first and last 5 axes 
are denoted by $\theta_{2}$ and $\theta_{3}$, respectively, and 
the azimuthal angles by $\varphi_{j}^{(2)}$.  The rotations
through $2\pi/3$ around those three-fold axes are denoted by $R_{j}$, 
$1\leq j \leq 10$. The two-fold axes join the midpoints of two 
opposite edges. The polar and azimuthal angles of the first, next 
and last 5 axes are $\theta_{4}$, $\varphi_{j}^{(1)}$, $\theta_{5}$, 
$\varphi_{j}^{(2)}$, $\pi$, and $\varphi_{j}^{(3)}$, respectively.
The rotations through $\pi$ around those two-fold axes are denoted 
by $S_{j}$, $1 \leq j \leq 15$. Those angles $\theta_{i}$ and 
$\varphi_{j}^{(i)}$ are given as follows:
$$\begin{array}{lll}
\tan \theta_{1}=2,~~~~  &\tan \theta_{2}=3-\sqrt{5}=2p^{2},~~~~ 
&\tan \theta_{3}=3+\sqrt{5}=2p^{-2},\\
\tan \theta_{4}=\left(\sqrt{5}-1\right)/2=p,~~~~
&\tan \theta_{5}=\left(\sqrt{5}+1\right)/2=p^{-1},~~~~& \\
\varphi_{j}^{(1)}=2(j-1)\pi/5, &\varphi_{j}^{(2)}=(2j-1)\pi/5,
&\varphi_{j}^{(3)}=(4j-3)\pi/10 , \\
p=\eta+\eta^{-1}, &p^{-1}=1+\eta+\eta^{-1}, &\eta=\exp(-i2\pi/5). 
\end{array} \eqno (3) $$

It is easy to see from Fig. 1 that 12 elements $E$, $S_{8}$,
$S_{12}$, $S_{1}$, $R_{6}^{\pm 1}$, $R_{2}^{\mp 1}$, $R_{4}^{\pm 1}$
and $R_{10}^{\mp 1}$ construct a subgroup $T$. 
Now, any element $R$ of {\bf I} can be expressed 
as a product of $T_{0}^{a}$ and an element 
$R_{6}^{b}S_{1}^{c}S_{12}^{d}$ of the subgroup $T$:
$$R=T_{0}^{a}R_{6}^{b}S_{1}^{c}S_{12}^{d}, \eqno (4) $$

Owing to the relations:
$$R_{6}=S_{1}T_{0}^{2}S_{1}T_{0}^{4},~~~~S_{12}=R_{6}^{2}S_{1}R_{6} ,
 \eqno (5) $$

\noindent
$T_{0}$ and $S_{1}$ are the generators of group {\bf I}. The
rank of {\bf I} is two. 

\vspace{5mm}
\noindent
{\bf III. IRREDUCIBLE BASES IN {\bf I} AND {\bf I}$_{h}$ 
GROUP SPACES}

\vspace{3mm}
It is convenient to choose the irreducible representations
of {\bf I} such that the representation matrices of one 
generator $T_{0}$ are diagonal. Assume that the bases $\Phi_{\mu \nu}$
in the {\bf I} group space are the eigenstates of 
left-action and right-action of $T_{0}$:
$$\begin{array}{ll}
T_{0}~\Phi_{\mu \nu}=\eta^{\mu}\Phi_{\mu \nu},~~~~~
&\Phi_{\mu \nu}~T_{0}=\eta^{\nu}\Phi_{\mu \nu}, \\
\eta=\exp(-i2\pi/5),~~~~~~ & \mu,~~\nu~~{\rm mod}~~5 ~.
\end{array} \eqno (6) $$

\noindent
The eigenstates can be easily calculated by the projection 
operator $P_{\mu}$ (see p.113 in Hamermesh 1962): 
$$\Phi_{\mu \nu}=c~P_{\mu}~R~P_{\nu},~~~~~
P_{\mu}=\displaystyle {1 \over 5} \sum_{\lambda=-2}^{2}~
\eta^{-\mu \lambda}~T_{0}^{\lambda}, \eqno (7) $$

\noindent
where $c$ is a normalization factor. The choice of the
group element $R$ in (7) will not affect the results except for
the factor $c$. In the following we choose $E$, $S_{11}$, 
$S_{5}$ and $S_{10}$ as the group element $R$, respectively,
and obtain four independent sets of bases $\Phi_{\mu \nu}^{(i)}$: 
$$\begin{array}{rl}
\Phi^{(1)}_{\mu \mu}&=~\left(E+\eta^{- \mu}T_{0}+\eta^{-2 \mu}T^{2}_{0}
+\eta^{2 \mu}T^{3}_{0}+\eta^{\mu} T^{4}_{0} \right)/\sqrt{5}~, \\
\Phi^{(2)}_{\mu \overline{\mu}}&=~\left(S_{11}+\eta^{-\mu}S_{14}+\eta^{-2 \mu}S_{12}
+\eta^{2 \mu}S_{15}+\eta^{\mu} S_{13} \right)/\sqrt{5}~, \\
\Phi^{(3)}_{\mu \nu}&=~\left\{\left(S_{5}+\eta^{-\mu}R^{2}_{5}
+\eta^{-2 \mu}T^{4}_{1}+\eta^{2 \mu}T_{4}+\eta^{\mu} R_{4} \right)\right.\\
&~~~+~\eta^{(\mu-\nu)}\left(S_{4}+\eta^{-\mu}R^{2}_{4}
+\eta^{-2 \mu}T^{4}_{5}+\eta^{2 \mu}T_{3}+\eta^{\mu} R_{3} \right)\\
&~~~+~\eta^{2(\mu-\nu)}\left(S_{3}+\eta^{-\mu}R^{2}_{3}
+\eta^{-2 \mu}T^{4}_{4}+\eta^{2 \mu}T_{2}+\eta^{\mu} R_{2} \right)\\
&~~~+~\eta^{-2(\mu-\nu)}\left(S_{2}+\eta^{-\mu}R^{2}_{2}
+\eta^{-2 \mu}T^{4}_{3}+\eta^{2 \mu}T_{1}+\eta^{\mu} R_{1} \right)\\
&~~~+~\eta^{-(\mu-\nu)}\left.\left(S_{1}+\eta^{-\mu}R^{2}_{1}
+\eta^{-2 \mu}T^{4}_{2}+\eta^{2 \mu}T_{5}+\eta^{\mu} R_{5} \right)
\right\}/5~, \\
\Phi^{(4)}_{\mu \nu}&=~\left\{\left(S_{10}+\eta^{-\mu}T^{3}_{1}
+\eta^{-2 \mu}R^{2}_{6}+\eta^{2 \mu}R_{9}+\eta^{\mu} T^{2}_{5} \right)\right.\\
&~~~+~\eta^{(\mu-\nu)}\left(S_{9}+\eta^{-\mu}T^{3}_{5}
+\eta^{-2 \mu}R^{2}_{10}+\eta^{2 \mu}R_{8}+\eta^{\mu} T^{2}_{4} \right)\\
&~~~+~\eta^{2(\mu-\nu)}\left(S_{8}+\eta^{-\mu}T^{3}_{4}
+\eta^{-2 \mu}R^{2}_{9}+\eta^{2 \mu}R_{7}+\eta^{\mu} T^{2}_{3} \right)\\
&~~~+~\eta^{-2(\mu-\nu)}\left(S_{7}+\eta^{-\mu}T^{3}_{3}
+\eta^{-2 \mu}R^{2}_{8}+\eta^{2 \mu}R_{6}+\eta^{\mu} T^{2}_{2} \right)\\
&~~~+~\eta^{-(\mu-\nu)}\left.\left(S_{6}+\eta^{-\mu}T^{3}_{2}
+\eta^{-2 \mu}R^{2}_{7}+\eta^{2 \mu}R_{10}+\eta^{\mu} T^{2}_{1} \right)
\right\}/5~, \end{array} \eqno (8) $$

\noindent
where and hereafter the subscript $\overline{\mu}$ denotes $-\mu$. Those
bases $\Phi_{\mu \nu}^{(i)}$ should be combined into the irreducible
bases $\psi_{\mu \nu}^{\Gamma}$ that belong to the given irreducible 
representation $\Gamma$. The combinations can be determined
from the condition that $\psi_{\mu \nu}^{\Gamma}$ should be the 
eigenstate of a class operator $W$, which was called CSCO-I
by Chen and Ping (1997). The eigenvalues $\alpha_{\Gamma}$ can be calculated 
from the characters in the irreducible representations $\Gamma$
(see (3-170) in Hamermesh 1962):
$$\begin{array}{ll}
W=\displaystyle \sum_{j=0}^{5}~\left(T_{j}+T_{j}^{4}\right),~~~~~
&W~\psi_{\mu \nu}^{\Gamma}=\psi_{\mu \nu}^{\Gamma}~W=
\alpha_{\Gamma}~\psi_{\mu \nu}^{\Gamma}, \\
\alpha_{A}=12,~~~~\alpha_{T_{1}}=4p^{-1},~~~~
&\alpha_{T_{2}}=-4p,~~~~\alpha_{G}=-3,~~~~
\alpha_{H}=0.
\end{array} \eqno (9) $$

\noindent
Now we calculate the matrix form of $W$ in the bases 
$\Phi_{\mu \nu}^{(i)}$, and diagonalize it. $\psi_{\mu \nu}^{\Gamma}$
are just the eigenvectors of the matrix form of $W$:
$$\psi_{\mu \nu}^{\Gamma}=N^{-1/2}~\displaystyle \sum_{i=1}^{4}~
C_{i}~\Phi_{\mu \nu}^{(i)} , \eqno (10) $$

\noindent
where $N$ is the normalization factor. In those bases
$\psi_{\mu \nu}^{\Gamma}$, the representation matrices are
diagonal with the diagonal elements $\eta^{\mu}$ (see (6)). 
In principle, each $\psi_{\mu \nu}^{\Gamma}$ contains a free 
phase, and the representation matrices of another generator 
$S_{1}$ depend upon the choice of phases.
We choose the phases such that the representation matrices
of $S_{1}$ are as follows:
$$\begin{array}{c}
D^{A}(S_{1})=1,~~~~~~~~~~
D^{T_{1}}(S_{1})=\displaystyle {1 \over \sqrt{5} }
\left( \begin{array}{ccc} -p^{-1} & -\sqrt{2} & -p \\
-\sqrt{2} & 1 & \sqrt{2} \\ -p & \sqrt{2} & -p^{-1} \end{array} \right),\\
\end{array} $$
$$\begin{array}{c}
D^{T_{2}}(S_{1})=\displaystyle {1 \over \sqrt{5} }
\left( \begin{array}{ccc} -p & \sqrt{2} & p^{-1} \\
\sqrt{2} & -1 & \sqrt{2} \\ p^{-1} & \sqrt{2} & -p \end{array} \right) ,~~~~
D^{G}(S_{1})=\displaystyle {1 \over \sqrt{5} }
\left( \begin{array}{cccc} -1 & -p & -p^{-1} & 1 \\
-p & 1 & -1 & -p^{-1} \\ -p^{-1} & -1 & 1 & -p \\
1 & -p^{-1} & -p & -1 \end{array} \right),  \\
D^{H}(S_{1})=\displaystyle {1 \over 5 }
\left( \begin{array}{ccccc} p^{-2} & 2p^{-1} & \sqrt{6} & 2p 
& p^{2} \\
2p^{-1} & p^{2} & -\sqrt{6} & -p^{-2} & -2p \\
 \sqrt{6} & -\sqrt{6} & -1 & \sqrt{6} & \sqrt{6} \\ 
2p & -p^{-2} & \sqrt{6} & p^{2} & -2p^{-1} \\ 
p^{2} & -2p & \sqrt{6} & -2p^{-1} & p^{-2} \end{array} \right) .  
\end{array} \eqno (11) $$

\noindent
where the row (column) indices $\mu$ of the irreducible 
representations $\Gamma$ are put in the following order: 
0 for $A$, 1, 0, and $\overline{1}$ for $T_{1}$, 2, 0, 
and $\overline{2}$ for $T_{2}$, 2, 1, $\overline{1}$, 
and $\overline{2}$ for $G$, and 2, 1, 0, $\overline{1}$, 
and $\overline{2}$ for $H$. The representation matrices 
of some irreducible representations of {\bf I} coincide 
with those in the subduced representations of $D^{\ell}$ of SO(3):
$$\begin{array}{c}
D^{0}(R)=D^{A}(R),~~~~D^{1}(R)=D^{T_{1}}(R),~~~~D^{2}(R)=D^{H}(R), \\
X^{-1}D^{3}(R)X=D^{T_{2}}(R)\oplus D^{G}(R),~~~~~R \in {\bf I}, \\
X=\left(\begin{array}{ccccccc} 0 & 0 & -\sqrt{2/5} & 0 & 0 & 0 
&\sqrt{3/5} \\ \sqrt{3/5} & 0 & 0 & -\sqrt{2/5} & 0 & 0 & 0 \\
0 & 0 & 0 & 0 & 1 & 0 & 0 \\ 0 & 1 & 0 & 0 & 0 & 0 & 0 \\
0 & 0 & 0 & 0 & 0 & 1 & 0 \\ 0 & 0 & \sqrt{3/5} & 0 & 0 & 0 &\sqrt{2/5} \\
\sqrt{2/5} & 0 & 0 & \sqrt{3/5} & 0 & 0 & 0 \end{array} \right) ,
\end{array} \eqno (12) $$

The normalization factors $N$ and combination coefficients 
$C_{i}$ in the expression (10) of $\psi_{\mu \nu}^{\Gamma}$
are listed in Table I. 

\begin{center}
\fbox{Table I}
\end{center}

The group {\bf I}$_{h}$ is the direct product of {\bf I} and
the inversion group $\left\{E, P\right\}$, where $P$ is the 
inversion operator. According to the parity, the irreducible 
representations of {\bf I}$_{h}$ are denoted as $\Gamma_{g}$ (even) and 
$\Gamma_{u}$ (odd) with the following irreducible bases: 
$$\psi_{\mu \nu}^{\Gamma_{g}}=2^{-1/2} \left(E+P\right)
\psi_{\mu \nu}^{\Gamma},~~~~~~~
\psi_{\mu \nu}^{\Gamma_{u}}=2^{-1/2} \left(E-P\right)
\psi_{\mu \nu}^{\Gamma}. \eqno (13) $$

Now we are in the position to construct the symmetry adapted bases 
(SAB). For a given polyatomic molecule with {\bf I}
or {\bf I}$_{h}$ symmetry, its vibrational states are described 
by the vibration quanta occupying in its bonds. Applying
the irreducible bases $\psi_{\mu \nu}^{\Gamma}$ to
the vibrational states, we obtain the SAB generally. The only 
problem is to determine the actions of group elements $R$
on the vibrational states according to the geometric
meaning of $R$. In fact, the action of $R$ only permutes, 
but does not change the vibration quanta. When some quanta 
are equal to each other, some SAB may be vanishing, or 
linearly dependent on other states. Let us give three examples to
explain the general method of calculating SAB.

{\bf Ex. 1}. The eigenvalues and eigenfunctions of
the H\"{u}ckel Hamiltonian for Carbon-60.

Deng and Yang (1992) have calculated this problem by
computer. Now we calculate the same problem in terms of 
the irreducible bases $\psi_{\mu \nu}^{\Gamma}$ even by
hand. It is easy to see from Fig. 1 in Deng and Yang (1992) that 
there is one-to-two correspondence between their states 
$|a,b,c\rangle$ and the group elements $R$ and $PR'$ of 
{\bf I}$_{h}$ in the following meaning:
$$R~|1,0,1\rangle=PR'~|1,0,1\rangle =|a,b,c\rangle,~~~~~
R ~~{\rm and}~~PR'~\longrightarrow ~|a,b,c\rangle  . \eqno (14) $$

\noindent
where $R\in $ {\bf I} and $R'\in $ {\bf I}. Introduce the 
new notation for the states:
$$|R\rangle=P~|R'\rangle=|PR'\rangle \equiv |a,b,c\rangle,~~~~~
P~|a,b,c\rangle=|\overline{a},b,c\rangle, \eqno (15) $$

\noindent
where the correspondence between $|a=1,b,c\rangle$ and the elements
$R$ and $PR'$ of {\bf I}$_{h}$ is listed as follows:
$$|R\rangle=|1,b,c\rangle,~~~~~
|R'\rangle=|\overline{1},b,c\rangle. \eqno (16) $$
\begin{center}
\begin{tabular}{c|cccccc}
$R(R')$  &$c=1$ &$c=2$ &$c=3$ &$c=4$ &$c=5$ &$c=6$ \\ \hline
$b=0$ & $E(S_{12})$ & $S_{1}(S_{8})$ & $R_{5}^{2}(T_{4}^{2})$ 
& $R_{1}(T_{3}^{3})$ & $T_{5}^{4}(R_{9})$ & $T_{2}(R_{7}^{2})$  \\
$b=1$ & $T_{0}(S_{15})$ & $R_{1}^{2}(T_{4}^{3})$ & $T_{1}^{4}(S_{9})$ 
& $S_{2}(R_{8}^{2})$ & $T_{3}(T_{5}^{2})$ & $R_{2}(R_{10})$  \\
$b=2$ & $T_{0}^{2}(S_{13})$ & $T_{2}^{4}(R_{9}^{2})$ & $T_{4}(T_{5}^{3})$ 
& $R_{2}^{2}(R_{6})$ & $R_{3}(S_{10})$ & $S_{3}(T_{1}^{2})$  \\
$b=3$ & $T_{0}^{3}(S_{11})$ & $T_{5}(R_{7})$ & $R_{4}(R_{10}^{2})$ 
& $T_{3}^{4}(T_{2}^{2})$ & $S_{4}(T_{1}^{3})$ & $R_{3}^{2}(S_{6})$  \\
$b=4$ & $T_{0}^{4}(S_{14})$ & $R_{5}(T_{3}^{2})$ & $S_{5}(R_{8})$ 
& $T_{1}(S_{7})$ & $R_{4}^{2}(R_{6}^{2})$ & $T_{4}^{4}(T_{2}^{3})$  
\end{tabular} 
\end{center}

Substituting (16) into (8), we obtain:
$$P~|\Phi_{\mu \mu}^{(1)}\rangle=
\eta^{2\mu}~|\Phi_{\mu \overline{\mu}}^{(2)}\rangle,~~~~
P~|\Phi_{\mu \nu}^{(3)}\rangle=
\eta^{2\mu-\nu}~|\Phi_{\mu \overline{\nu}}^{(4)}\rangle.
\eqno (17) $$

\noindent
Thus, some bases in (13) become vanishing or linearly dependent
on other bases. The independent bases are listed as follows:
$$\begin{array}{llll}
2^{-1/2}~|\psi^{A_{g}}_{00}\rangle=|\psi^{A}_{00}\rangle, ~~
&|\psi^{T_{1g}}_{\mu 1}\rangle=|\psi^{T_{1g}}_{\mu \overline{1}}\rangle,
&|\psi^{T_{1u}}_{\mu 1}\rangle=-|\psi^{T_{1u}}_{\mu \overline{1}}\rangle,
&2^{-1/2}~|\psi^{T_{1u}}_{\mu 0}\rangle=|\psi^{T_{1}}_{\mu 0}\rangle,\\
|\psi^{T_{2g}}_{\mu 2}\rangle=-|\psi^{T_{2g}}_{\mu \overline{2}}\rangle,~
&|\psi^{T_{2u}}_{\mu 2}\rangle=|\psi^{T_{2u}}_{\mu \overline{2}}\rangle,~
&2^{-1/2}~|\psi^{T_{2u}}_{\mu 0}\rangle=|\psi^{T_{2}}_{\mu 0}\rangle,~~
&|\psi^{G_{g}}_{\mu 2}\rangle=|\psi^{G_{g}}_{\mu \overline{2}}\rangle,\\
|\psi^{G_{g}}_{\mu 1}\rangle=|\psi^{G_{g}}_{\mu \overline{1}}\rangle,
&|\psi^{G_{u}}_{\mu 2}\rangle=-|\psi^{G_{u}}_{\mu \overline{2}}\rangle,
&|\psi^{G_{u}}_{\mu 1}\rangle=-|\psi^{G_{u}}_{\mu \overline{1}}\rangle,
&|\psi^{H_{g}}_{\mu 2}\rangle=|\psi^{H_{g}}_{\mu \overline{2}}\rangle,\\
|\psi^{H_{g}}_{\mu 1}\rangle=-|\psi^{H_{g}}_{\mu \overline{1}}\rangle,~~~
&2^{-1/2}~|\psi^{H_{g}}_{\mu 0}\rangle=|\psi^{H}_{\mu 0}\rangle,~~~
&|\psi^{H_{u}}_{\mu 2}\rangle=-|\psi^{H_{u}}_{\mu \overline{2}}\rangle,~~~
&|\psi^{H_{u}}_{\mu 1}\rangle=|\psi^{H_{u}}_{\mu \overline{1}}\rangle.
\end{array} \eqno (18) $$

\noindent
where an additional normalization factor $2^{-1/2}$ has to be
introduced when
$$|\psi^{\Gamma}_{\mu 0}\rangle=P |\psi^{\Gamma}_{\mu 0}\rangle
~~~~{\rm or}~~~~
|\psi^{\Gamma}_{\mu 0}\rangle=-P |\psi^{\Gamma}_{\mu 0}\rangle. $$

There are 90 bonds that are divided into two types (Deng and Yang 1992). 
The $hp$ bonds separate a hexagon from a pentagon, and the others are
called the $hh$ bonds. Following the notation in Deng and Yang (1992), the
H\"{u}ckel interaction of the 60 $hp$ bonds are equal to
$-\alpha$, and that of the 30 $hh$ bonds equal to $(\alpha-2)$. 
Now, since the states of $C_{60}$ are denoted by the elements 
$R$ of {\bf I}, the action of Hamiltonian on the states can be 
written from (16) and the Figure in Deng and Yang (1992), for example:
$$\begin{array}{l}
H~|E\rangle=-\alpha~|T_{0}\rangle-\alpha~|T_{0}^{4}\rangle
+(\alpha-2)~|S_{1}\rangle , \\
H~|T_{0}\rangle=-\alpha~|E\rangle-\alpha~|T_{0}^{2}\rangle
+(\alpha-2)~|R_{1}^{2}\rangle , \\
H~|T_{0}^{4}\rangle=-\alpha~|E\rangle-\alpha~|T_{0}^{3}\rangle
+(\alpha-2)~|R_{5}\rangle , \\
H~|S_{1}\rangle=-\alpha~|R_{1}\rangle-\alpha~|R_{5}^{2}\rangle
+(\alpha-2)~|E\rangle . \end{array} \eqno (19) $$

\noindent
We are only interested in the properties of $H$ acting on $|E\rangle$
and vice versa.
The matrix of the Hamiltonian in the irreducible bases (18)
is a hermitian and block one, which can be calculated by the 
standard method of group theory (Hamermesh 1962). For example, there 
are two sets of bases $|\psi_{\mu 1}^{T_{1u}}\rangle$ and 
$|\psi_{\mu 0}^{T_{1}}\rangle$ for the representation $T_{1u}$:
$$\begin{array}{rl}
|\psi_{11}^{T_{1u}}\rangle&=~8^{-1/2}\left\{|\Phi_{11}^{(1)}\rangle
-\eta^{2}|\Phi_{1\overline{1}}^{(2)}\rangle
-p^{-1}|\Phi_{11}^{(3)}\rangle
+p\eta^{2}|\Phi_{1\overline{1}}^{(3)}\rangle
-p|\Phi_{11}^{(4)}\rangle
+p^{-1}\eta|\Phi_{1\overline{1}}^{(4)}\rangle \right\} \\
&=~(200)^{-1/2}~\left\{\sqrt{5}\left(|E\rangle
+\eta^{-1}|T_{0}\rangle+\eta|T_{0}^{4}\rangle\right)+(-p^{-1}+p)
|S_{1}\rangle+\cdots \right\} , \\[1mm]
|\psi_{10}^{T_{1}}\rangle &=
~2^{-1/2}\left\{-\eta |\Phi_{10}^{(3)}\rangle
+\eta^{-2}|\Phi_{10}^{(4)}\rangle \right\} 
=(50)^{-1/2}~
\left\{-|S_{1}\rangle+\cdots \right\} , \\[1mm]
|\psi_{01}^{T_{1u}}\rangle &=~
2^{-1}\left\{-\eta^{-1}|\Phi_{01}^{(3)}\rangle
-\eta|\Phi_{0\overline{1}}^{(3)}\rangle+\eta^{2}|\Phi_{01}^{(4)}\rangle
+\eta^{-2}|\Phi_{0\overline{1}}^{(4)}\rangle \right\} 
=(10)^{-1}~\left\{-2|S_{1}\rangle+\cdots \right\} , \\[1mm]
|\psi_{00}^{T_{1}}\rangle&=~
2^{-1}\left\{|\Phi_{00}^{(1)}\rangle
-|\Phi_{00}^{(2)}\rangle +|\Phi_{00}^{(3)}\rangle
-|\Phi_{00}^{(4)}\rangle \right\} \\
&=~(10)^{-1}~\left\{\sqrt{5}\left(|E\rangle
+|T_{0}\rangle+|T_{0}^{4}\rangle\right)+|S_{1}\rangle+\cdots \right\} ,
\end{array} \eqno (20) $$

\noindent
where we only list the terms of $|E\rangle$, $|T_{0}\rangle$,
$|T_{0}^{4}\rangle$, and $|S_{1}\rangle$, which are relevant 
to the calculation. Comparing the coefficients of the term 
$|E\rangle$ on both sides of the following equations:
$$\begin{array}{lll}
H~|\psi_{11}^{T_{1u}}\rangle=H_{11}~|\psi_{11}^{T_{1u}}\rangle
+H_{01}~|\psi_{10}^{T_{1}}\rangle ,\\
H~|\psi_{10}^{T_{1}}\rangle=H_{10}~|\psi_{11}^{T_{1u}}\rangle
+H_{00}~|\psi_{10}^{T_{1}}\rangle ,\\
H~|\psi_{01}^{T_{1u}}\rangle=H_{11}~|\psi_{01}^{T_{1u}}\rangle
+H_{01}~|\psi_{00}^{T_{1}}\rangle , \\
H~|\psi_{00}^{T_{1}}\rangle=H_{10}~|\psi_{01}^{T_{1u}}\rangle
+H_{00}~|\psi_{00}^{T_{1}}\rangle ,
\end{array} \eqno (21) $$

\noindent
we obtain the submatrix of $H$ for the representation $T_{1u}$,
which is of two dimensions:
$$\begin{array}{l}
H^{T_{1u}}=\displaystyle { 1 \over 2\sqrt{5} }
\left(\begin{array}{cc} -\alpha(7-\sqrt{5})+4 & -4(\alpha-2) \\
-4(\alpha-2) & -2\alpha(2\sqrt{5}-1)-4 \end{array} \right) , \\
E^{T_{1u}}=-\alpha\left(3+\sqrt{5}\right)/4\pm \displaystyle 
{1 \over 4} \left\{18\alpha^{2}
\left(3-\sqrt{5}\right)-16\alpha\left(5-\sqrt{5}\right)+64\right\}^{1/2} .
\end{array}  \eqno (22) $$

\noindent
The advantage of this method is that the eigenfunctions of
the Hamiltonian are able to be obtained simultaneously.

In the same way we can easily calculate the submatrices of the 
H\"{u}ckel Hamiltonian for other representations: {\small
$$\begin{array}{l}
H^{A_{g}}=-\alpha-2,~~~~
H^{T_{1g}}=-\alpha\left(\sqrt{5}+1\right)/2+2,~~~~
H^{T_{2g}}=\alpha\left(\sqrt{5}-1\right)/2+2, \\
H^{T_{2u}}=\displaystyle { 1 \over 2\sqrt{5} }
\left(\begin{array}{cc} \alpha(7+\sqrt{5})-4 & 4(\alpha-2) \\
4(\alpha-2) & -2\alpha(2\sqrt{5}+1)+4 \end{array} \right),~~~
H^{G_{g}}=\left(\begin{array}{cc} \alpha(\sqrt{5}+1)/2 & -(\alpha-2) \\
-(\alpha-2) & -\alpha(\sqrt{5}-1)/2 \end{array} \right),\\
H^{G_{u}}=\displaystyle { 1 \over 2\sqrt{5} }
\left(\begin{array}{cc} \alpha(\sqrt{5}+1)+8 & 2(\alpha-2) \\
2(\alpha-2) & \alpha(\sqrt{5}-1)-8 \end{array} \right), ~~~
H^{H_{u}}=\displaystyle { 1 \over 2\sqrt{5} }
\left(\begin{array}{cc} \alpha(7+\sqrt{5})-4 & 4(\alpha-2) \\
4(\alpha-2) & -\alpha(7-\sqrt{5})+4 \end{array} \right),\\
\end{array} $$
$$\begin{array}{l}
H^{H_{g}}=\displaystyle {1 \over 10}
\left(\begin{array}{ccc} \alpha(5\sqrt{5}+11)-12 & 4(\alpha-2)
& 4\sqrt{3}(\alpha-2) \\  4(\alpha-2)
& -\alpha(5\sqrt{5}-11)-12 & -4\sqrt{3}(\alpha-2) \\
4\sqrt{3}(\alpha-2) & -4\sqrt{3}(\alpha-2) 
& -22 \alpha+4 \end{array} \right).
\end{array} \eqno (23) $$ }

\noindent
The dimensions of the submatrices $H^{\Gamma}$ are one or two
except for $H^{H_{g}}$, so that the energy levels of the H\"{u}ckel 
Hamiltonian are able to be calculated by hand: 
$$\begin{array}{l}
E^{A_{g}}=-\alpha-2,~~~~~ 
E^{T_{1g}}=-\alpha \left(\sqrt{5}+1\right)/2+2,~~~~~
E^{T_{2g}}=\alpha \left(\sqrt{5}-1\right)/2+2,\\
E^{T_{2u}}=-\alpha\left(3-\sqrt{5}\right)/4\pm \displaystyle 
{1 \over 4} \left\{18\alpha^{2}
\left(3+\sqrt{5}\right)-16\alpha\left(5+\sqrt{5}\right)+64\right\}^{1/2} ,\\
E^{G_{g}}=\alpha/2 \pm 2^{-1} \left(9\alpha^{2}-16\alpha
+16\right)^{1/2} ,~~~~~
E^{G_{u}}=\alpha/2 \pm 2^{-1}\left(\alpha^{2}
+16\right)^{1/2},\\
E^{H_{u}}=\alpha/2 \pm 2^{-1} \left(13\alpha^{2}
-24\alpha  +16\right)^{1/2} ,
\end{array} \eqno (24) $$  

\noindent
and $E^{H_{g}}$ is the root of the following equation:
$$x^{3}+2x^{2}+x\left(-6\alpha^{2}+8\alpha-4\right)
+\left(\alpha^{3}-12\alpha^{2}+16\alpha-8\right)=0.
\eqno (25) $$

\noindent
The results coincide with that given in Deng and Yang (1992) except for a
misprint in (32) of Deng and Yang (1992) ($Q_{3'-}$ and $Q_{3-}$
should be switched, but Fig. 2 in Deng and Yang (1992) is correct.) 

{\bf Ex. 2}. The submatrices of the H\"{u}ckel
Hamiltonian for carbon-240.

It can be seen from Fig. 1 of Chou and Yang (1993) that, in comparison 
with each atom of Carbon-60, Carbon-240 contains three more 
carbon atoms distributed symmetrically around that carbon 
atom of Carbon-60. In addition to $(a,b,c)$, 
we introduce a new index $\lambda$ to identify those four 
carbon atoms. The carbon atom in the center is labeled by 
$\lambda=1$, the carbon on the hexagon labeled by 2, and 
the carbons on the two neighbor pentagons labeled by 3 
and 4, respectively. Each carbon atom corresponds to a state, 
denoted by $|a,b,c,\lambda \rangle$, or by a group element 
$R$ and $\lambda$ in terms of the generalized notation in (15):
$$\begin{array}{c}
|R,\lambda \rangle=|PR',\sigma(\lambda) \rangle=|a,b,c,\lambda \rangle,
~~~~P~|a,b,c,\lambda \rangle =|\overline{a},b,c,\sigma(\lambda)\rangle, \\
\sigma(1)=1,~~~~~\sigma(2)=2,~~~~~\sigma(3)=4,~~~~~\sigma(4)=3.
\end{array} \eqno (26) $$

\noindent
From (8) we have:
$$\begin{array}{c}
P|\Phi^{(1)}_{\mu \mu},\lambda \rangle=
\eta^{2\mu}~|\Phi^{(2)}_{\mu \overline{\mu}}, \sigma(\lambda) \rangle,~~~~~
P|\Phi^{(3)}_{\mu \nu},\lambda \rangle=
\eta^{2\mu-\nu}~|\Phi^{(4)}_{\mu \overline{\nu}}, \sigma(\lambda) \rangle.
\end{array} \eqno (27) $$

Following (10), (13) and (18), we are able to combine
$|\Phi^{(i)}_{\mu \nu},\lambda \rangle$
into the SAB $|\Gamma, \mu, \tau \rangle$. For example, 
we have three independent SAB for $\Gamma=A_{1g}$ and one
for $\Gamma=A_{1u}$ : 
$$\begin{array}{rl}
|A_{1g},0,1\rangle &=~|\psi^{A}_{00},1\rangle
=60^{-1/2}\left\{|E,1\rangle+\cdots \right\}, \\
|A_{1g},0,2\rangle&=~|\psi^{A}_{00},2\rangle
=60^{-1/2}\left\{|E,2\rangle+|T_{0},2\rangle+|T_{0}^{4},2\rangle
+\cdots \right\}, \\
|A_{1g},0,3\rangle&=~2^{-1/2}\left\{|\psi^{A}_{00},3\rangle
+|\psi^{A}_{00},4\rangle \right\} \\
&=~120^{-1/2}\left\{|E,3\rangle+|E,4\rangle+|T_{0}^{4},3\rangle
+|T_{0},4\rangle+|S_{1},3\rangle+|S_{1},4\rangle+\cdots \right\}, \\
|A_{1u},0,1\rangle&=~2^{-1/2}\left\{|\psi^{A}_{00},3\rangle
-|\psi^{A}_{00},4\rangle \right\} \\
&=~120^{-1/2}\left\{|E,3\rangle-|E,4\rangle+|T_{0}^{4},3\rangle
-|T_{0},4\rangle+|S_{1},3\rangle-|S_{1},4\rangle+\cdots \right\}, \\
\end{array} \eqno (28) $$

In the bases $|R,\lambda\rangle$, the matrix of the H\"{u}ckel 
Hamiltonian $H$ is hermitian. We are only 
interested in the property of $H$ acting on the states
$|E,\lambda \rangle$ and vice versa. For bond arrangement (a) we have:
$$\begin{array}{l}
H~|E,1\rangle=-\alpha~|E,3\rangle-\alpha~|E,4\rangle
+(\alpha-2)~|E,2\rangle \\
H~|E,2\rangle=-\alpha~|T_{0},2\rangle-\alpha~|T_{0}^{4},2\rangle
+(\alpha-2)~|E,1\rangle \\
H~|E,3\rangle=-\alpha~|E,1\rangle-\alpha~|S_{1},4\rangle
+(\alpha-2)~|T_{0},4\rangle \\
H~|E,4\rangle=-\alpha~|E,1\rangle-\alpha~|S_{1},3\rangle
+(\alpha-2)~|T_{0}^{4},3\rangle \\
\end{array} \eqno (29) $$

\noindent
and for bond arrangement (b):
$$\begin{array}{l}
H~|E,1\rangle=-\alpha~|E,3\rangle-\alpha~|E,4\rangle
+(\alpha-2)~|E,2\rangle \\
H~|E,2\rangle=-\alpha~|T_{0},2\rangle-\alpha~|T_{0}^{4},2\rangle
+(\alpha-2)~|E,1\rangle \\
H~|E,3\rangle=-\alpha~|E,1\rangle+(\alpha-2)~|S_{1},4\rangle
-\alpha~|T_{0},4\rangle \\
H~|E,4\rangle=-\alpha~|E,1\rangle+(\alpha-2)~|S_{1},3\rangle
-\alpha~|T_{0}^{4},3\rangle \\
\end{array} \eqno (30) $$

\noindent
Therefore, in the expansions of (28) we only need list 10 
relevant states:
$$\begin{array}{lllll}
|E,1\rangle,~~&|E,2\rangle,~~&|E,3\rangle,~~&|E,4\rangle ,~~
&|T_{0},2\rangle,\\
|T_{0},4\rangle, &|T_{0}^{4},2\rangle, &|T_{0}^{4},3\rangle,
&|S_{1},3\rangle, &|S_{1},4\rangle
\end{array} \eqno (31) $$

\noindent
From (28), (29) and (30) we obtain the submatrices of the H\"{u}ckel
Hamiltonian for the representations $A_{1g}$ and $A_{1u}$:
$$H^{A_{1g}}(a)=H^{A_{1g}}(b)=\left(\begin{array}{ccc}
0 & \alpha-2 & -\sqrt{2}\alpha \\
\alpha-2 & -2\alpha & 0 \\ -\sqrt{2} \alpha & 0 & -2   
\end{array} \right),~~~~H^{A_{1u}}(a)=H^{A_{1u}}(b)=2. \eqno (32) $$

\noindent
In the cases $A_{1g}$ and $A_{1u}$, the submatrices of $H$ are same 
for both bond arrangements (a) and (b). There are more SAB belonging
to other irreducible representations. However, the calculations
are still simple enough to complete by hand. In the following we
list the independent SAB for each irreducible representation
and the nonvanishing matrix elements of the Hamiltonian .
$$\begin{array}{ll}
|T_{1g},\mu , \lambda \rangle=2^{-1/2}\left\{
|\psi_{\mu 1}^{T_{1}},\lambda \rangle
+|\psi_{\mu \overline{1}}^{T_{1}},\sigma(\lambda) \rangle \right\}, ~~
&|T_{2g},\mu , \lambda \rangle=2^{-1/2}\left\{
|\psi_{\mu 2}^{T_{2}},\lambda \rangle
-|\psi_{\mu \overline{2}}^{T_{2}},\sigma(\lambda) \rangle \right\} ,\\
|T_{1g},\mu , 5\rangle=2^{-1/2}\left\{|\psi_{\mu 0}^{T_{1}},3\rangle
-|\psi_{\mu 0}^{T_{1}},4 \rangle \right\} ,
&|T_{2g},\mu , 5\rangle=2^{-1/2}\left\{|\psi_{\mu 0}^{T_{2}},3\rangle
-|\psi_{\mu 0}^{T_{2}},4 \rangle \right\} ,
\end{array} \eqno (33) $$
$$\begin{array}{ll}
|T_{1u},\mu , \lambda \rangle=2^{-1/2}\left\{
|\psi_{\mu 1}^{T_{1}},\lambda \rangle
-|\psi_{\mu \overline{1}}^{T_{1}},\sigma(\lambda) \rangle \right\} ,~~
&|T_{2u},\mu , \lambda \rangle=2^{-1/2}\left\{
|\psi_{\mu 2}^{T_{2}},\lambda \rangle
+|\psi_{\mu \overline{2}}^{T_{2}},\sigma(\lambda) \rangle \right\} ,\\
|T_{1u},\mu , 5 \rangle=2^{-1/2}\left\{
|\psi_{\mu 0}^{T_{1}},3 \rangle
+|\psi_{\mu 0}^{T_{1}},4 \rangle \right\}, 
&|T_{2u},\mu , 5 \rangle=2^{-1/2}\left\{
|\psi_{\mu 0}^{T_{2}},3 \rangle
+|\psi_{\mu 0}^{T_{2}},4 \rangle \right\}, \\
|T_{1u},\mu , 6 \rangle=|\psi_{\mu 0}^{T_{1}},1 \rangle ,
&|T_{2u},\mu , 6 \rangle=|\psi_{\mu 0}^{T_{2}},1 \rangle ,\\
|T_{1u},\mu , 7 \rangle=|\psi_{\mu 0}^{T_{1}},2 \rangle , 
&|T_{2u},\mu , 7 \rangle=|\psi_{\mu 0}^{T_{2}},2 \rangle , \\
\end{array} \eqno (34) $$

\noindent
where $\lambda$ runs from 1 to 4, and $\sigma(\lambda)$
was given in (26). Since the submatrices of the Hamiltonian
are all hermitian, we only list the nonvanishing matrix 
elements in the up-triangle 
part (the row index is not larger than the column index).
$$\begin{array}{l}
H^{T_{1g}}(a)_{22}=H^{T_{1g}}(b)_{22}=H^{T_{1u}}(a)_{22}=H^{T_{1u}}(b)_{22}=
-\alpha p, \\
H^{T_{1g}}(a)_{33}=H^{T_{1g}}(a)_{44}=-H^{T_{1u}}(a)_{33}
=-H^{T_{1u}}(a)_{44}= \alpha p /\sqrt{5},\\
H^{T_{1g}}(b)_{33}=H^{T_{1g}}(b)_{44}=-H^{T_{1u}}(b)_{33}
=-H^{T_{1u}}(b)_{44}=-(\alpha-2) p/\sqrt{5}, \\
\end{array} $$
$$\begin{array}{l}
H^{T_{1g}}(a)_{55}=-H^{T_{1u}}(a)_{55}=-(\alpha -2)+\alpha/\sqrt{5}, \\
H^{T_{1g}}(b)_{55}=-H^{T_{1u}}(b)_{55}=\alpha -(\alpha-2)/\sqrt{5}, \\
H^{T_{1u}}(a)_{77}=H^{T_{1u}}(b)_{77}=-2\alpha, \\
H^{T_{1g}}(a)_{12}=H^{T_{1g}}(b)_{12}
=H^{T_{1u}}(a)_{12}=H^{T_{1u}}(b)_{12}
=H^{T_{1u}}(a)_{67}=H^{T_{1u}}(b)_{67}=\alpha-2, \\
H^{T_{1g}}(a)_{13}=H^{T_{1g}}(a)_{14}=H^{T_{1g}}(b)_{13}=H^{T_{1g}}(b)_{14}
=H^{T_{1u}}(a)_{13}=H^{T_{1u}}(a)_{14}\\
~~~=H^{T_{1u}}(b)_{13}=H^{T_{1u}}(b)_{14}=-\alpha, \\
H^{T_{1g}}(a)_{34}=H^{T_{1u}}(a)_{34}
=(\alpha-2)\eta^{-1}+\alpha p^{-1}/\sqrt{5}, \\
H^{T_{1g}}(b)_{34}=H^{T_{1u}}(b)_{34}
=-\alpha \eta^{-1}-(\alpha-2) p^{-1}/\sqrt{5}, \\
-H^{T_{1g}}(a)_{35}=H^{T_{1g}}(a)_{45}=H^{T_{1u}}(a)_{35}
=H^{T_{1u}}(a)_{45}=\alpha \sqrt{2/5}, \\
-H^{T_{1g}}(b)_{35}=H^{T_{1g}}(b)_{45}=H^{T_{1u}}(b)_{35}
=H^{T_{1u}}(b)_{45}=-(\alpha-2) \sqrt{2/5}, \\
H^{T_{1u}}(a)_{56}=H^{T_{1u}}(b)_{56}=-\alpha \sqrt{2}. \\
\end{array} \eqno (35) $$

\noindent
After the replacement of $\sqrt{5}$ by $-\sqrt{5}$ from
the submatrices for $T_{1}$ representation, we obtain
those for $T_{2}$. 

For the representations $G$ and $H$ we have
$$\begin{array}{ll}
|G_{g},\mu , \lambda \rangle=2^{-1/2}\left\{
|\psi_{\mu 2}^{G},\lambda \rangle
+|\psi_{\mu \overline{2}}^{G},\sigma(\lambda) \rangle \right\} ,~~
&|G_{u},\mu , \lambda \rangle=2^{-1/2}\left\{
|\psi_{\mu 2}^{G},\lambda \rangle
-|\psi_{\mu \overline{2}}^{G},\sigma(\lambda) \rangle \right\} ,\\
|G_{g},\mu , 4+\lambda \rangle=2^{-1/2}\left\{
|\psi_{\mu 1}^{G},\lambda \rangle
+|\psi_{\mu \overline{1}}^{G},\sigma(\lambda) \rangle \right\} ,
&|G_{u},\mu , 4+\lambda \rangle=2^{-1/2}\left\{
|\psi_{\mu 1}^{G},\lambda \rangle
-|\psi_{\mu \overline{1}}^{G},\sigma(\lambda) \rangle \right\} .
\end{array} \eqno (36) $$
$$\begin{array}{l}
H^{G_{g}}(a)_{22}=H^{G_{g}}(b)_{22}=H^{G_{u}}(a)_{22}=H^{G_{u}}(b)_{22}=
\alpha p^{-1}, \\
H^{G_{g}}(a)_{66}=H^{G_{g}}(b)_{66}=H^{G_{u}}(a)_{66}=H^{G_{u}}(b)_{66}=
-\alpha p, \\
H^{G_{g}}(a)_{33}=H^{G_{g}}(a)_{44}=-H^{G_{g}}(a)_{77}=-H^{G_{g}}(a)_{88}\\
~~~=-H^{G_{u}}(a)_{33}=-H^{G_{u}}(a)_{44}=H^{G_{u}}(a)_{77}=H^{G_{u}}(a)_{88}=
-\alpha /\sqrt{5}, \\
H^{G_{g}}(b)_{33}=H^{G_{g}}(b)_{44}=-H^{G_{g}}(b)_{77}=-H^{G_{g}}(b)_{88}\\
~~~=-H^{G_{u}}(b)_{33}=-H^{G_{u}}(b)_{44}=H^{G_{u}}(b)_{77}=H^{G_{u}}(b)_{88}=
(\alpha-2) /\sqrt{5}, \\
H^{G_{g}}(a)_{12}=H^{G_{g}}(a)_{56}=H^{G_{u}}(a)_{12}=H^{G_{u}}(a)_{56}\\
~~~=H^{G_{g}}(b)_{12}=H^{G_{g}}(b)_{56}=H^{G_{u}}(b)_{12}=H^{G_{u}}(b)_{56}=
\alpha-2, \\
H^{G_{g}}(a)_{13}=H^{G_{g}}(a)_{14}=H^{G_{g}}(a)_{57}=H^{G_{g}}(a)_{58}=
H^{G_{u}}(a)_{13}=H^{G_{u}}(a)_{14}\\
~~~=H^{G_{u}}(a)_{57}=H^{G_{u}}(a)_{58}
=H^{G_{g}}(b)_{13}=H^{G_{g}}(b)_{14}=H^{G_{g}}(b)_{57}=H^{G_{g}}(b)_{58}\\
~~~=H^{G_{u}}(b)_{13}=H^{G_{u}}(b)_{14}=H^{G_{u}}(b)_{57}=H^{G_{u}}(b)_{58}=
-\alpha, \\
H^{G_{g}}(a)_{34}=H^{G_{u}}(a)_{34}=(\alpha-2)\eta^{-2}+\alpha/\sqrt{5}, \\
H^{G_{g}}(b)_{34}=H^{G_{u}}(b)_{34}=-\alpha \eta^{-2}-(\alpha-2)/\sqrt{5}, \\
H^{G_{g}}(a)_{78}=H^{G_{u}}(a)_{78}=(\alpha-2)\eta^{-1}-\alpha/\sqrt{5}, \\
H^{G_{g}}(b)_{78}=H^{G_{u}}(b)_{78}=-\alpha \eta^{-1}+(\alpha-2)/\sqrt{5}, \\
H^{G_{g}}(a)_{37}=H^{G_{g}}(a)_{48}=-H^{G_{u}}(a)_{37}=-H^{G_{u}}(a)_{48}=
\alpha p^{-1}/\sqrt{5}, \\
H^{G_{g}}(b)_{37}=H^{G_{g}}(b)_{48}=-H^{G_{u}}(b)_{37}=-H^{G_{u}}(b)_{48}=
-(\alpha-2) p^{-1}/\sqrt{5}, \\
H^{G_{g}}(a)_{38}=H^{G_{g}}(a)_{47}=H^{G_{u}}(a)_{38}=H^{G_{u}}(a)_{47}=
\alpha p/\sqrt{5}, \\
H^{G_{g}}(b)_{38}=H^{G_{g}}(b)_{47}=H^{G_{u}}(b)_{38}=H^{G_{u}}(b)_{47}=
-(\alpha-2) p/\sqrt{5}. 
\end{array} \eqno (37) $$
$$\begin{array}{ll}
|H_{g},\mu , \lambda \rangle=2^{-1/2}\left\{
|\psi_{\mu 2}^{H},\lambda \rangle
+|\psi_{\mu \overline{2}}^{H},\sigma(\lambda) \rangle \right\} ,~~
&|H_{u},\mu , \lambda \rangle=2^{-1/2}\left\{
|\psi_{\mu 2}^{H},\lambda \rangle
-|\psi_{\mu \overline{2}}^{H},\sigma(\lambda) \rangle \right\} ,\\
|H_{g},\mu , 4+\lambda \rangle=2^{-1/2}\left\{
|\psi_{\mu 1}^{H},\lambda \rangle
-|\psi_{\mu \overline{1}}^{H},\sigma(\lambda) \rangle \right\} , 
&|H_{u},\mu , 4+\lambda \rangle=2^{-1/2}\left\{
|\psi_{\mu 1}^{H},\lambda \rangle
+|\psi_{\mu \overline{1}}^{H},\sigma(\lambda) \rangle \right\} , \\
|H_{g},\mu , 9 \rangle=2^{-1/2}\left\{
|\psi_{\mu 0}^{H}, 3 \rangle
+|\psi_{\mu 0}^{H}, 4 \rangle \right\}, 
&|H_{u},\mu , 9 \rangle=2^{-1/2}\left\{
|\psi_{\mu 0}^{H}, 3 \rangle
-|\psi_{\mu 0}^{H}, 4 \rangle \right\}, \\
|H_{g},\mu , 10 \rangle=|\psi_{\mu 0}^{H},1 \rangle , &\\
|H_{g},\mu , 11 \rangle=|\psi_{\mu 0}^{H},2 \rangle . &\\
\end{array} \eqno (38) $$
$$\begin{array}{l}
H^{H_{g}}(a)_{22}=H^{H_{g}}(b)_{22}=H^{H_{u}}(a)_{22}=H^{H_{u}}(b)_{22}=
\alpha p^{-1}, \\
H^{H_{g}}(a)_{66}=H^{H_{g}}(b)_{66}=H^{H_{u}}(a)_{66}=H^{H_{u}}(b)_{66}=
-\alpha p, \\
H^{H_{g}}(a)_{33}=H^{H_{g}}(a)_{44}=-H^{H_{u}}(a)_{33}=-H^{H_{u}}(a)_{44}
=-\alpha p^{2}/5, \\
H^{H_{g}}(b)_{33}=H^{H_{g}}(b)_{44}=-H^{H_{u}}(b)_{33}=-H^{H_{u}}(b)_{44}
=(\alpha-2) p^{2}/5, \\
H^{H_{g}}(a)_{77}=H^{H_{g}}(a)_{88}=-H^{H_{u}}(a)_{77}=-H^{H_{u}}(a)_{88}
=-\alpha p^{-2}/5, \\
H^{H_{g}}(b)_{77}=H^{H_{g}}(b)_{88}=-H^{H_{u}}(b)_{77}=-H^{H_{u}}(b)_{88}
=(\alpha-2) p^{-2}/5, \\
H^{H_{g}}(a)_{99}=-H^{H_{u}}(a)_{99}=(\alpha-2)+\alpha/5, \\
H^{H_{g}}(b)_{99}=-H^{H_{u}}(b)_{99}=-\alpha-(\alpha-2)/5, \\
H^{H_{g}}(a)_{11,11}=H^{H_{g}}(b)_{11,11}=-2\alpha , \\
H^{H_{g}}(a)_{12}=H^{H_{g}}(a)_{56}=H^{H_{g}}(a)_{10,11}=
H^{H_{u}}(a)_{12}=H^{H_{u}}(a)_{56}\\
~~~=H^{H_{g}}(b)_{12}=H^{H_{g}}(b)_{56}=H^{H_{g}}(b)_{10,11}=
H^{H_{u}}(b)_{12}=H^{H_{u}}(b)_{56}=\alpha-2,\\
H^{H_{g}}(a)_{13}=H^{H_{g}}(a)_{14}=H^{H_{g}}(a)_{57}=H^{H_{g}}(a)_{58}
=H^{H_{u}}(a)_{13}=H^{H_{u}}(a)_{14}\\
~~~=H^{H_{u}}(a)_{57}=H^{H_{u}}(a)_{58}=H^{H_{g}}(b)_{13}
=H^{H_{g}}(b)_{14}=H^{H_{g}}(b)_{57}=H^{H_{g}}(b)_{58}\\
~~~=H^{H_{u}}(b)_{13}=H^{H_{u}}(b)_{14}=H^{H_{u}}(b)_{57}=H^{H_{u}}(b)_{58}
=-\alpha,\\
H^{H_{g}}(a)_{34}=H^{H_{u}}(a)_{34}=(\alpha-2)\eta^{-2}-\alpha p^{-2}/5,\\
H^{H_{g}}(b)_{34}=H^{H_{u}}(b)_{34}=-\alpha \eta^{-2}+(\alpha-2) p^{-2}/5,\\
H^{H_{g}}(a)_{78}=H^{H_{u}}(a)_{78}=(\alpha-2)\eta^{-1}-\alpha p^{2}/5,\\
H^{H_{g}}(b)_{78}=H^{H_{u}}(b)_{78}=-\alpha \eta^{-1}+(\alpha-2) p^{2}/5,\\
H^{H_{g}}(a)_{37}=H^{H_{g}}(a)_{48}=
-H^{H_{u}}(a)_{37}=-H^{H_{u}}(a)_{48}=2\alpha p/5,\\
H^{H_{g}}(b)_{37}=H^{H_{g}}(b)_{48}=
-H^{H_{u}}(b)_{37}=-H^{H_{u}}(b)_{48}=-2(\alpha-2) p/5,\\
H^{H_{g}}(a)_{38}=H^{H_{g}}(a)_{47}=
H^{H_{u}}(a)_{38}=H^{H_{u}}(a)_{47}=-2\alpha p^{-1}/5,\\
H^{H_{g}}(b)_{38}=H^{H_{g}}(b)_{47}=
H^{H_{u}}(b)_{38}=H^{H_{u}}(b)_{47}=2(\alpha-2) p^{-1}/5,\\
H^{H_{g}}(a)_{39}=H^{H_{g}}(a)_{49}=-H^{H_{g}}(a)_{79}=-H^{H_{g}}(a)_{89}\\
~~~=-H^{H_{u}}(a)_{39}=H^{H_{u}}(a)_{49}=H^{H_{u}}(a)_{79}=-H^{H_{u}}(a)_{89}
=-\alpha \sqrt{6}/5,\\
H^{H_{g}}(b)_{39}=H^{H_{g}}(b)_{49}=-H^{H_{g}}(b)_{79}=-H^{H_{g}}(b)_{89}\\
~~~=-H^{H_{u}}(b)_{39}=H^{H_{u}}(b)_{49}=H^{H_{u}}(b)_{79}=-H^{H_{u}}(b)_{89}
=(\alpha-2) \sqrt{6}/5,\\
H^{H_{g}}(a)_{9,10}=H^{H_{g}}(b)_{9,10}=-\alpha \sqrt{2}. \\
\end{array} \eqno (39) $$

The secular equations can be calculated by a standard program
in Mathematica, and coincide with those given in Chou and Yang (1993)
except for one dropped zero there. The coefficient of the term 
$\lambda^{6}\alpha^{5}$ in $Q_{5+}$ is 100, not 10.

{\bf Ex. 3}. The symmetry adapted bases of $B_{12}H_{12}$.

A state in $B_{12}H_{12}$ is described by the vibration
quanta in the 12 bonds. Those numbers of the vibration 
quanta are denoted by $n_{j}$ and $m_{j}$ for the 
bonds $OA_{j}$ and $OB_{j}$, $0\leq j \leq 5$, respectively. 
Applying the irreducible bases $\psi_{\mu \nu}^{\Gamma}$
on the states, we obtain the SAB as follows:
$$\psi_{\mu \nu}^{\Gamma}~|n_{0}n_{1}n_{2}n_{3}n_{4}n_{5}
m_{0}m_{1}m_{2}m_{3}m_{4}m_{5} \rangle , \eqno (40) $$

\noindent
where the action of a group element $R$ of {\bf I} on the
state can be calculated from the definition of $R$
and from Fig. 1. For example, 
$$\begin{array}{l}
A_{0},A_{1},A_{2},A_{3},A_{4},A_{5} ~\stackrel{T_{0}} {\longrightarrow}~
A_{0},A_{2},A_{3},A_{4},A_{5},A_{1}, \\
A_{0},A_{1},A_{2},A_{3},A_{4},A_{5}
~\stackrel{S_{11}} {\longrightarrow}~
B_{0},B_{4},B_{3},B_{2},B_{1},B_{5}, \\
A_{0},A_{1},A_{2},A_{3},A_{4},A_{5}
~\stackrel{S_{5}} {\longrightarrow}~
A_{5},A_{4},B_{2},B_{3},A_{1},A_{0}, \\
A_{0},A_{1},A_{2},A_{3},A_{4},A_{5}
~\stackrel{S_{10}} {\longrightarrow}~
B_{3},A_{5},B_{2},B_{0},B_{4},A_{1} . 
\end{array} $$

\noindent
Under the applications of $T_{0}$, $S_{11}$, $S_{5}$ and $S_{10}$, 
the state $|n_{0}n_{1}n_{2}n_{3}n_{4}n_{5}
m_{0}m_{1}m_{2}m_{3}m_{4}m_{5} \rangle  $ becomes:
$$\begin{array}{ll}
T_{0}:~~~&|n_{0}n_{5}n_{1}n_{2}n_{3}n_{4} 
m_{0}m_{5}m_{1}m_{2}m_{3}m_{4}\rangle, \\
S_{11}:~~~&|m_{0}m_{4}m_{3}m_{2}m_{1}m_{5}
n_{0}n_{4}n_{3}n_{2}n_{1}n_{5} \rangle, \\
S_{5}:~~~&|n_{5}n_{4}m_{2}m_{3}n_{1}n_{0}
m_{5}m_{4}n_{2}n_{3}m_{1}m_{0} \rangle , \\
S_{10}:~~~&|m_{3}n_{5}m_{2}m_{0}m_{4}n_{1}
n_{3}m_{5}n_{2}n_{0}n_{4}m_{1} \rangle . \\
\end{array} \eqno (41) $$

\noindent
When 12 quanta are all different from each other, we obtain
60 SAB that are divided into 16 sets with given irreducible 
representations. If some quanta are equal to each other, 
the number of the independent sets may decrease. Since the 
dimensions of the representations are less than 60 for the 
seven important cases discussed in Chen and Ping (1997), those 
representations were called non-regular (Chen and Ping 1997).

\vspace{5mm}
\noindent
{\bf IV. CONCLUSION}

\vspace{3mm}
The symmetry adapted bases are very useful in calculating
the eigenvalues and eigenstates of a Hamiltonian with
given symmetry. From the irreducible bases in the group space 
of the symmetry group of the system, the SAB can be calculated
generally and simply. This is a standard method in group 
theory (Hamermesh 1962), and widely used in the problems of vibrations of a 
polyatomic molecule (Lemus and Frank 1994, Ma {\it et al}. 1996, 
Chen {\it et al}. 1996). The explicit form of the 
irreducible bases of {\bf I} group space are useful in the future
calculations for the molecules with {\bf I} and {\bf I}$_{h}$
symmetry.

\vspace{5mm}
\noindent
{\bf ACKNOWLEDGMENTS} 

The authors would like to thank Professor 
Jin-Quan Chen for useful discussion. This work was supported by 
the National Natural Science Foundation of China and Grant No. 
LWTZ-1298 of Chinese Academy of Sciences.

\vspace{5mm}
\noindent
{\bf REFERENCES}

\noindent
Altmann, S. L., and Herzig, P. (1994). {\it Point-Group Theory
Tables}, Oxford University Press, Oxford.

\noindent
Balasubramanian, K. (1996). {\it Chemical Physics Letters}, {\bf 260}, 476.

\noindent
Brown, W. B. (1987). {\it Chemical Physics Letters}, {\bf 136}, 128 and
{\bf 139}, 612.

\noindent
Chen, J. Q., Klein, A., and Ping, J. L. (1996). {\it Journal of 
Mathematical Physics }, {\bf 37}, 2400. 

\noindent
Chen, J. Q., and Ping, J. L. (1997). 
{\it Journal of Mathematical Physics}, {\bf 38}, 387.

\noindent
Chou, T. T., and Yang, C. N. (1993). {\it Physics Letters A}, 
{\bf 183}, 221.

\noindent
Clougherty, D. P., and Gorman, J. P. (1996). {\it Chemical 
Physics Letters}, {\bf 251}, 353.

\noindent
Deng, Y., and Yang, C. N. (1992). {\it Physics Letters A}, 
{\bf 170}, 116.

\noindent
Doye J. P. K., and Wales, D. J. (1996). {\it Chemical 
Physics Letters}, {\bf 262}, 167.

\noindent
Fowler P. W., and Ceulemans, A. (1985). {\it Molecular Physics}, 
{\bf 54}, 767.

\noindent
Friedberg, H. R., and Lee, T. D. (1992). {\it Physical Review B}, 
{\bf 46}, 14150.

\noindent
Giannozzi, P., and Baroni, S. (1994). {\it Journal of Chemical Physics}, 
{\bf 100}, 8537.

\noindent
Golding, R. M. (1973). {\it Molecular Physics}, {\bf 26}, 661.

\noindent
Gunnarsson, O., Handschuh, H., Bechthold, P. S.,
Kessler, B., Gantefoer, G., and Eberhardt, W. (1995). 
{\it Physical Review Letters}, {\bf 74}, 1875.

\noindent
Hamermesh, M. (1962). {\it Group Theory and its 
Application to Physical Problems}, Addison-Wesley Pub. Co.
Reading. 

\noindent
Huffman, D. R. (1991). {\it Physics Today}, Nov. 22.

\noindent
Kroto, H. W. (1988). {\it Science}, {\bf 242}, 1139.

\noindent
Kroto, H. W., Heath, J. R., O'Brien, S. C., 
Curl, R. F., and Smalley, R. E. (1985). {\it Nature}, {\bf 318}, 162.

\noindent
Lemus R., and Frank, A. (1994). {\it Journal of Chemical Physics}, 
{\bf 101}, 8321.

\noindent
Liu, F., Ping, J. L., and Chen, J. Q. (1990).
{\it Journal of Mathematical Physics}, {\bf 31}, 1065.

\noindent
Lomont, J. S. (1959). {\it Applications of Finite Groups}, 
Academic Press, New York, p.32 and p.312. 

\noindent
Ma, Z. Q., Hou, X. W., and Xie, M. (1996). {\it Physical 
Review A}, {\bf 53}, 2173.

\noindent
Mart\'{i}nez-Torres, E., L\'{o}pez-Gonz\'{a}lez, J. J.,
Fern\'{a}ndez-G\'{o}mez, M., Brendsdal, E., and Cyvin, S. J. (1996).
{\it Chemical Physics Letters}, {\bf 253}, 32.

\noindent
McLellan, A. G. (1961). {\it Journal of Chemical Physics}, {\bf 34}, 1350.

\noindent
Negri, F., and Orlandi, G. (1996). {\it Journal of Physics B}, 
{\bf 29}, 5049. 

\noindent
Olthof, E. H. T., van der Avoired, A., and Wormer, P. E. S. (1996). 
{\it Journal of Chemical Physics}, {\bf 104}, 832.

\noindent
Pennis, E. (1991). {\it Science News}, {\bf 140}, 120.

\noindent
Pooler, D. R. (1980). {\it Journal of Physics A}, {\bf 13}, 1197.

\noindent
Rohlfing, E. A., Cox, D. M., and Kaldor, A. (1984)
{\it Journal of Chemical Physics}, {\bf 81}, 3322.

\noindent
Schettino, V., Salvi, P. R., Bini, R., and Cardini, G. (1994).
{\it Journal of Chemical Physics}, {\bf 101}, 11079.

\noindent
Tang, A. C., and Huang, F. Q. (1997). {\it International Journal 
of Quantum Chemistry}, {\bf 63}, 367.

\noindent
Tang, A. C., Huang, F. Q., and Liu, R. Z. (1996). {\it Physical 
Review B}, {\bf 53}, 7442.

\noindent
Varga, F., Nemes, L., and Watson, J. K. G. (1996). {\it Journal 
of Physics B}, {\bf 29}, 5043.

\noindent
Wang, Z., Day, P., and Pachter, R. (1996). {\it Chemical 
Physics Letters}, {\bf 248}, 121.

\noindent
Weeks, D. E., and Harter, W. G. (1989). 
{\it Journal of Chemical Physics}, {\bf 90}, 4744.
\end{document}